\documentclass[prl,twocolumn,showpacs]{revtex4}
\usepackage{txfonts}
\usepackage{graphicx,amssymb,mathrsfs,bm,color,times}

\newcommand{\be}{\begin{equation}}
\newcommand{\ee}{\end{equation}}
\newcommand{\bea}{\begin{eqnarray}}
\newcommand{\eea}{\end{eqnarray}}
\newcommand{\bsube}{\begin{subequations}}
\newcommand{\esube}{\end{subequations}}

\newcommand{\Eq}[1]{Eq.\,(\ref{#1})}

\begin{document}
\title {Quantum Dynamics of Mesoscopic Driven Duffing Oscillators}

\author{Lingzhen Guo }
\affiliation{Department of Physics, Beijing Normal University,
Beijing 100875, China}
\author{Zhigang Zheng}
\affiliation{Department of Physics, Beijing Normal University,
Beijing 100875, China}
\author{Xin-Qi Li\footnote{Corresponding author: xqli@red.semi.ac.cn}}
\affiliation{Department of Physics, Beijing Normal University,
Beijing 100875, China}

\date{\today}

\begin{abstract}
We investigate the nonlinear dynamics of a mesoscopic driven
Duffing oscillator in a quantum regime. In terms of Wigner
function, we identify the nature of state near the bifurcation
point, and analyze the transient process which reveals two
distinct stages of quenching and escape. The rate process in the
escape stage allows us to extract the transition rate, which
displays perfect scaling behavior with the driving distance to the
bifurcation point. We numerically determine the scaling exponent,
compare it with existing result, and propose open questions to be
resolved.
\end{abstract}

 \maketitle


A broad class of physical systems such as Josephson junction,
trapped electron or ion, and nano-mechanical oscillator, can be
well described by the Duffing oscillator under proper conditions.
One of the most profound features of a driven Duffing oscillator
(DDO) is the dynamical bifurcation. Near the bifurcation point,
the oscillator state is highly sensitive to perturbation. This
property can be exploited for applications such as sensing device,
amplifier, and logic device. Most recently, for instance, the
superconductor circuit based on Josephson junction has been
exploited for quantum measurement of superconducting qubits
~\cite{Siddiqi207002,Siddiqi027005,Devoret014524,Mooij119}. This
device is termed as Josephson bifurcation amplifier (JBA), holding
advantages such as fast speed, high sensitivity, low backaction,
and absence of on-chip dissipation.

Despite that the classical bifurcation of DDO is well-known, the
quantum dynamics in the bistable region and near the bifurcation
point has been a new and significant subject in the past
years~\cite{Katz07,Pea06,Dykman042108,Dykman011101,MDQT,low-lying}.
This new trend is motivated mostly by the advent of approaching
the quantum regime of nano-mechanical oscillators, as well as the
bifurcation-based quantum measurement devices. For instance, the
quantum signature in the bistable region of a DDO was proposed,
based on simulating a Lindblad-type master equation and comparing
the Wigner function with classical probability distribution in
phase space~\cite{Katz07}. In terms of amplitude and phase
responses to the driving frequency, quantum behaviors of DDO such
as resonant tunneling and photon-assisted tunneling were also
discussed~\cite{Pea06}. Moreover, in
Ref.~\cite{Dykman042108,Dykman011101,MDQT}, switching rate between
the bistable states near the bifurcation point, due to quantum
and/or thermal fluctuations, is estimated by means of the WKB
theory or semiclassical methods such as the
mean-first-passage-time approach.

In this letter we consider a {\it mesoscopic} DDO, with about more
than ten levels that is in between the quantum few-level and the
classical dense-level (or continuum) limits. 
In this regime, the quantum effect is apparently significant.
However, at the same time, how the DDO's nonlinearity manifests
itself is unclear and of interest, since the few-level (e.g. 2- or
3-level) system should have no such behaviors as bistability and
bifurcation. Our present study will demonstrate the existence of
bistable region, characterize the quantum nature of the states,
and investigate the quantum transition near the classical
bifurcation point which displays {\it perfect} and {\it new}
scaling behavior with the driving strength.

{\it Model and Method}.--- The Duffing oscillator in the presence
of driving is described by the Hamiltonian
\begin{equation}\label{DDO}
\hat{H}_0(t)=\frac{\hat{p}^2}{2m}+\frac{1}{2}m\Omega^2\hat{x}^2
-\gamma\hat{x}^4+F(t)\hat{x}.
\end{equation}
For the JBA setup, $F(t)=F_{0}(e^{i\nu t}+e^{-i\nu t})$ describes
the microwave driving. Other parameters are related to the JBA
circuit quantities as: $m=(\hbar/2e)^2C,\
\Omega=\sqrt{2eI_{c}/(\hbar C)}, \ F_{0}=\hbar I/(2e),
\mathrm{and} \ \gamma=m\Omega^2/24$; with $C$ the capacitance of
the Josephson junction, $I_{c}$ the critical current, and $I$ the
driving current. In this context, $x$ denotes the phase difference
across the Josephson junction.

In addition, the Duffing oscillator is affected by environment,
which together with the coupling can be modelled as $
\hat{H}_E=\sum_{i}[m_i\omega_i^2\hat{x_i}^2/2+\hat{p}_i^2/2m_i]-
\hat{x}\sum_{i}\lambda_i\hat{x}_i+\hat{x}^2
\sum_i\lambda_i^2/(2m_i\omega_i^2). $ Typically, the spectral
density of the bath,
$J(\omega)=\pi\sum_{i}\lambda_{i}^2\delta(\omega-\omega_{i})
/(2m_i\omega_i)$, in Ohmic case reads
$J(\omega)=m\kappa\omega\mathrm{exp}(-\omega/\omega_c)$, with
$\kappa$ the friction coefficient, and $\omega_{c}$ the high
frequency cutoff. For later use, we also introduce
$\hat{b}=\sum_i\lambda_i\hat{b}_i/{\sqrt{2}}$, with
$\hat{b}_i=(m_i\omega_i\hat{x}_i+i\hat{p}_i)
/{\sqrt{2m_i\hbar\omega_i}}$.

In the weak coupling limit to the environment and under Markovian
approximation, the dissipative dynamics of the DDO is governed by
the quantum master equation (see Ref.~\cite{yanyijing} for more
details)
\begin{equation}\label{1}
 \dot{\rho}(t)=-\frac{i}{\hbar}[\hat{H}(t),\rho(t)]
 -\frac{1}{\hbar^2}\{[\hat{x},\hat{Q}\rho(t)]+\mathrm{H.c.}\}.
\end{equation}
Here, $\rho(t)$ is the reduced density matrix of the oscillator;
$\hat{H}(t)=\hat{H}_0(t)+\hat{x}^2m\kappa\omega_c/\pi$, and
$\hat{Q}=[C(\mathcal{-L})+\tilde{C}(\mathcal{-L})]\hat{x}/2$. The
Liouvillian $\mathcal{L}$ is defined through its action on an
arbitrary operator $\hat{O}$ as: $ \mathcal{L}\hat{O} \equiv
\hbar^{-1} [\hat{H}(t)-F(t)\hat{x},\hat{O}] $. The superoperators
$C(\mathcal{L})$ and $\tilde{C}(\mathcal{L})$ are the Fourier
transform of the bath correlation functions: $
C(\mathcal{L})=\int_{-\infty}^{+\infty} dtC(t)e^{i\mathcal{L}t}$,
and $
\tilde{C}(\mathcal{L})=\int_{-\infty}^{+\infty}dt\tilde{C}(t)
e^{i\mathcal{L}t} $. The correlators $C(t)$ and $\tilde{C}(t)$ are
defined by
$C(t)=\mathrm{Tr}_E[\hat{b}^\dagger(t)\hat{b}(0)\rho_E]$, and
$\tilde{C}(t)=\mathrm{Tr}_E [\hat{b}(t)\hat{b}^\dagger(0)\rho_E]
$, where $\rho_E$ is the thermal-equilibrium density operator of
the environment.

{\it Qualitative Considerations}.--- In the absence of driving,
the Duffing oscillator described by~\Eq{DDO} has only finite
number of bound states. This can be seen from the potential
profile, $V(x)=m\Omega^2x^2/2-\gamma x^4$, which defines a single
well with identical barrier height $V_0=m^2\Omega^4/(16\gamma)$ at
$x=\pm\sqrt{m\Omega^2/(4\gamma)}$. As a rough estimate, the number
of bound states is the ratio of $V_0$ and $\hbar\Omega$, which
gives $ N=m^2\Omega^4/(16\gamma
\hbar\Omega)=m\Omega/(16\hbar\tilde{\gamma})
=\aleph/(16\tilde{\gamma}). $ We will see later that $\aleph\equiv
m\Omega/\hbar$ defined here is an important characteristic
quantity. We also introduced a reduced nonlinear coefficient,
$\tilde{\gamma}=\gamma/(m\Omega^2)$. In our model,
$\gamma=m\Omega^2/24$, so approximately the number of bound state
is $3\aleph/2$. 
In the experiment of Ref.~\cite{Siddiqi207002}, $\aleph \simeq
366$, which implies a classical DDO. In the present work, we
consider a {\it mesoscopic} regime, by assuming possible
parameters $I_c=39\mathrm{nA},\ C=0.91\mathrm{pF},\
\kappa=0.01\Omega$, and $\omega_c=10\Omega$. Accordingly,
$\aleph\simeq 12$.

Under proper conditions~\cite{Devoret014524}, the DDO exhibits the
most profound phenomenon known as {\it bifurcation}. 
To determine the bifurcation point, we present an analysis in the
{\it rotating phase space}. Starting with the Hamiltonian
$H_0(t)$, we introduce a unitary transformation,
$\hat{U}=\exp[i\nu t(\hat{a}^\dagger\hat{a})]$, where $\hat{a}$ is
the annihilation operator of the Duffing oscillator. 
Under the rotating wave approximation (RWA), we obtain~\cite{MDQT}
\begin{eqnarray}\label{42}
\hat{H}_s^{(\delta)}
&=\left(\frac{\hat{p}^2}{2\tilde{m}}+\frac{1}{2}\tilde{m}
\tilde{\Omega}^2\hat{x}^2\right)-\frac{6\gamma}{4\tilde{m}^2
\tilde{\Omega}^4}\left(\frac{\hat{p}^2}{2\tilde{m}}+\frac{1}{2}
\tilde{m}\tilde{\Omega}^2\hat{x}^2\right)^2& \nonumber\\
&+F_0\hat{x},&
\end{eqnarray}
where $\delta=1-\nu/\Omega$, $\tilde{m}=m/\delta$, and
$\tilde{\Omega}=\Omega \delta$. In phase space, the extremal
points of $\hat{H}_s^{(\delta)}$ satisfies
\begin{eqnarray} \label{cubic_equation}
p=0, \ \ \ \ \ \
x^3-\frac{2\tilde{m}\tilde{\Omega}^2}{3\gamma}x-\frac{2F_0}{3\gamma}=0.
\end{eqnarray}
For this cubic equation, the discriminant reads $
\Delta=[-2F_0/(3\gamma\times 2)]^2
  +[-2\tilde{m}\tilde{\Omega}^2/(3\gamma\times 3)]^3  $.
If and only if $\Delta<0$, there exist three different real roots.
This implies
\begin{eqnarray}\label{Fc-2}
F_0&<&\frac{2}{9}(\frac{2\tilde{m}^3\tilde{\Omega}^6}{\gamma})^{1/2}
=\frac{8\sqrt{3}m\Omega^2\delta^{3/2}}{9}=F_c  .
\end{eqnarray}
This result, based on the existence condition of multiple steady
states, coincides with that from the singularity
analysis~\cite{Devoret014524}. The advantage of the present method
is its applicability to more general situation, e.g., the
mesoscopic case under present study, in which we will see that the
{\it singular bifurcation} is absent. Nevertheless, in this work
we still refer to the estimated $F_c$ of \Eq{Fc-2} as {\it
bifurcation point} for description convenience, particularly as a
reference value for the driving strength.

Moreover, still classically, to realize the bifurcation it was
found in Ref.~\cite{Devoret014524} that the driving frequency
$\nu$ should be lower than $\Omega$ and satisfy  $\delta
> \sqrt{3}/(2Q)$, where $Q=\Omega/\kappa$. 
Based on the above analysis, here we can further set a upper limit
to the detuning. For $F_0<F_c$, \Eq{cubic_equation} has three
different real roots, with the largest one as
$X_{max}=2[F_c/(3\gamma)]^{1/3}\cos\theta/3$, where
$\theta=\arctan\sqrt{(F_c/F_0)^2-1}$. Obviously, this largest
amplitude should not overcome the potential barrier of the Duffing
oscillator. This consideration leads to the following inequality:
$$
2(\frac{F_c}{3\gamma})^{1/3}<\sqrt{m\Omega^2/(4\gamma)}\Longrightarrow
\  \delta<\frac{9}{64(3\tilde{\gamma})^{1/3}},\ \
\tilde{\gamma}=\frac{\gamma}{m\Omega^2} .
$$
In our model, $\gamma=m\Omega^2/24$. So we get
$\delta<18/64\thickapprox 0.28$. This is not the optimized value.
More accurate consideration can result in even smaller upper
limit.

\begin{figure}
\center
\includegraphics[scale=0.7]{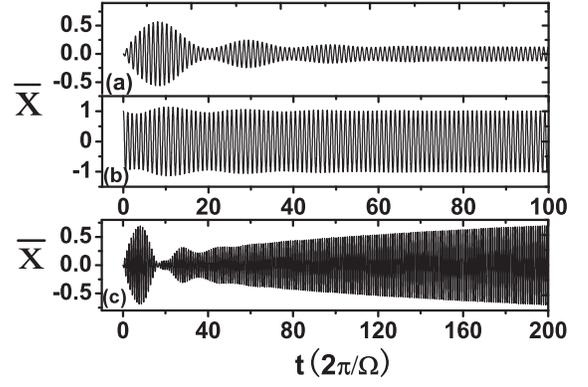}
\caption{ (a) and (b): Bistable feature visualized by the
transient behavior of $\bar{x}(t)$ =$\mathrm{Tr}[\hat{x}\rho(t)]$.
Parameters: driving strength $F_0=0.8 F_c$, and driving frequency
$\nu=0.94\Omega$. 
(c): Two successive stages (i.e. quenching and escape) towards the
LAS, for driving near the critical point as exemplified here by
$F_0=0.95F_c$. }
\end{figure}

\begin{figure}
 \center
 \includegraphics[scale=0.9]{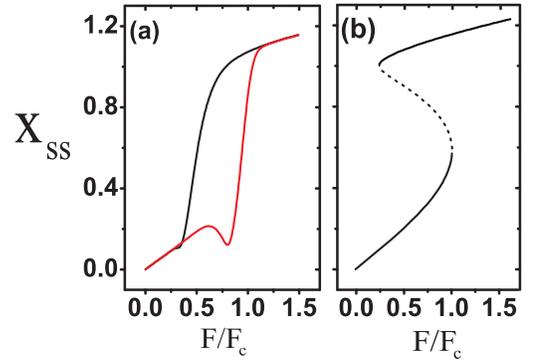}
 \caption{
(a) Phase diagram of the oscillation amplitude of stable state
against the driving strength. The red and black curves are for
initial states of Gaussian wavepackets centered at $x=0$ and
$x=1$, respectively. 
(b) Classical counterpart of (a), showing sharp bifurcation
behavior. }
\end{figure}

{\it Quantum Dynamics near the Bifurcation Point}.--- Based on a
direct simulation of the master equation, we show in Fig.\ 1 the
evolution of $\bar{x}(t)=\mathrm{Tr}[\hat{x}\rho(t)]$. 
First, in Fig.\ 1(a) and (b), we demonstrate the {\it bistable}
nature, for driving strength $F_0=0.8 F_c$ as an example. We
consider two initial conditions: the ground state, and a coherent
state centered at $\bar{x}=1$. Indeed, for the mesoscopic DDO,
here we find {\it quantum mechanically} that the steady state does
exhibit bistable behavior, say, depending on the initial
condition, it arrives at either a small amplitude state (SAS), or
a large amplitude state (LAS). 
Nevertheless, as we will understand later in the following study,
the results in Fig.\ 1(a) and (b) are not the {\it fundamental}
SAS and LAS, but their mixture with different population
probabilities depending on the initial conditions. 
For driving not very close to $F_c$, the ``steady state"
population is formed in relatively short time, and later the
fluctuation-induced transitions between the fundamental SAS and
LAS are negligibly weak.

In contrast, as shown in Fig.\ 1(c), for driving closer to $F_c$
(e.g. $F_0=0.95 F_c$) we find that the entire process contains two
distinct stages, say, a (fast) {\it quenching} stage, and a
successive (slow) {\it escape} stage. 
In the quenching stage, the oscillator rapidly arrives at the SAS.
Then, it is followed by a rate process (transition) to the LAS,
which may be termed as the Kramers {\it escape}
process~\cite{Hanggi}. Conventionally the escape is caused by
thermal fluctuations, as described by for instance the
mean-first-passage-time approach or Fokker-Planck
equation~\cite{Dykman011101,MDQT}. 
In Ref.~\cite{MDQT}, quantum-fluctuation induced transition was
also investigated, by using the WKB method. 
The advantage of our present numerical simulation allows to
formulate a way to extract the transition rate under more general
conditions, say, in the presence of both thermal and quantum
fluctuations, going beyond the existing results in limiting cases.
This will be detailed in latter analysis.

In Fig.\ 2(a) we extract data from numerical simulation as shown
in Fig.\ 1 to plot the {\it phase diagram}, say, the oscillation
amplitude of steady state against the driving strength, which
shows the desirable {\it hysteresis} behavior. 
However, compared to its classical counterpart as schematically
shown in Fig.\ 2(b), two differences should be remarked. {\it (i)}
The singularity of transition from the bistable region to the
single LAS or SAS disappears, although the transition point, i.e.,
$F_c=(8\sqrt{3}m\Omega^2\delta^{3/2})/9$, is approximately
preserved. 
The basic reason for this {\it gradual} transition behavior is
that for the present {\it mesoscopic} DDO, the stable state is a
statistical mixture of the SAS and LAS in the absence of noise
(i.e. thermal and quantum fluctuations). {\it (ii)} 
A dip appears in the red curve. Our numerical simulation shows
that the time-dependent oscillations of the SAS and LAS are out of
phase (i.e. with a phase difference about $\pi$), and the SAS and
LAS themselves depend on the driving strength {\it nonlinearly}.
As a consequence of interplay of these two factors, the {\it dip}
is formed as we observed.

\begin{figure}
 \center
 \includegraphics[scale=0.6]{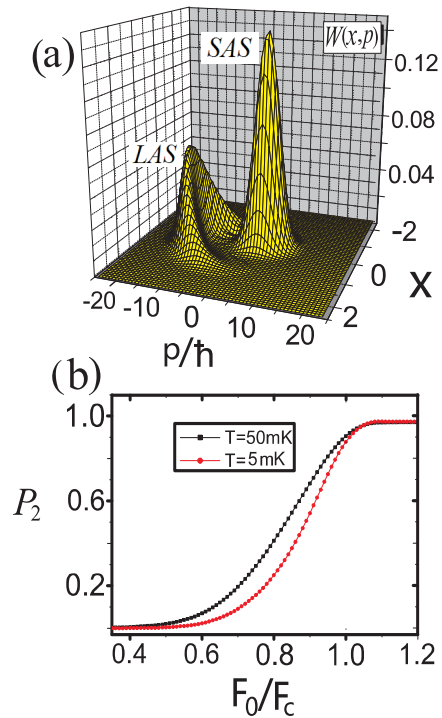}
 \caption{
(a) Wigner function at $t=160 \ast (2\pi/\Omega)$, for driving
$F_0=0.9F_c$ and starting with the ground state. The two separated
wavepackets correspond to the SAS and LAS, and the SAS is nearly a
perfect coherent state. 
(b) Occupation probability of the LAS ($P_2$) against the driving
strength. In the region near $F_c$, $P_2$ shows exponential
dependence of the driving strength. }
\end{figure}

Let us proceed our further analysis with the help of the Wigner
function, which is defined as: $ W(x,p,t)
=1/(\pi\hbar)\int_{-\infty}^{+\infty}\langle
x+x'|\rho(t)|x-x'\rangle\exp(-i2px'/\hbar)dx'$. The Wigner
function is widely used in broad context of physics, with an
intuitive interpretation of probability in the phase space. 
In Fig.\ 3(a) we show the Wigner function of the oscillator at
time $t=160 \ast (2\pi/\Omega)$, for driving strength
$F_0=0.9F_c$, and with the ground state as the initial condition.
Time dependently, the Wigner function is in fact rotating with the
driving frequency in phase space, along the classical trajectory
but with additional diffusion because of the thermal and/or
quantum fluctuations.

In the transient process, after certain duration time to be
discussed below, the oscillator can be well described by a mixed
state, formally as
\begin{equation}
 W(x,p,t)=P_1(t)W_S(x,p,t)+P_2(t) W_L(x,p,t).
\end{equation}
Here, $W_S(x,p,t)$ and $W_L(x,p,t)$ are, respectively, the Wigner
functions of the {\it intrinsic} SAS and LAS associated with the
given driving strength, but not the {\it averaged} ones shown in
Fig.\ 2(a). 
In essence, the SAS and LAS are two limit cycles, or attractors,
each with its own attraction basin~\cite{basin,Siddiqi207002}. 
$P_1(t)$ and $P_2(t)$ are the occupation probabilities of the SAS
and LAS. In Fig.\ 3(b) we plot $P_2(t)$ at $t=160 \ast
(2\pi/\Omega)$ versus the driving strength, for two temperatures.

Closer inspection indicates that $W_S(x,p,t)$ is quantum
mechanically a pure state, i.e., $\mathrm{Tr}\rho_{SAS}^2 \simeq
1$, where $\rho_{SAS}$ is the density matrix of the SAS. Moreover,
it is nearly a perfect coherent state. 
This result can be understood as follows. For a harmonic
oscillator under the interplay of driving and dissipation, such as
the optical cavity field under excitation and photon-loss, the
steady state is exactly a coherent state~\cite{Knight}. Then, for
the present {\it nonlinear} Duffing oscillator, since the SAS is
not far from the oscillator origin, it is thus approximately
governed by a harmonic oscillation. 
For LAS, however, which is far from the origin, nonlinearity is
prominent, which makes $W_L(x,p,t)$ not at all a coherent state,
but a mixed state with partial coherence.

\begin{figure}
 \center
 \includegraphics[scale=0.7]{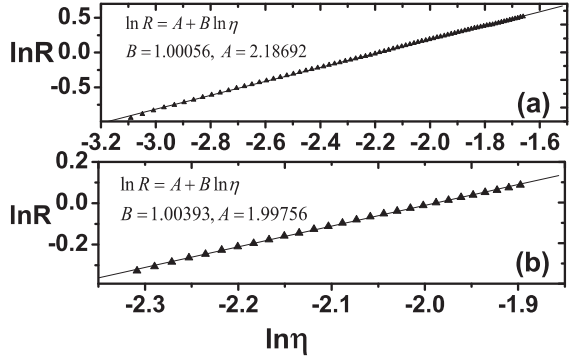}
 \caption{ Perfect scaling behavior hidden in the transition rate
with the driving distance to the critical point, say, $\eta =
F_c^2-F_0^2$. The triangles stand for data from numerical
simulation, and the straight lines are linear fits. The result
shows that $R \varpropto\eta^{\alpha}$, and $\alpha\simeq 1$.
Parameters: $\kappa=0.01$; $T$=5mK in (a), and $T$=50mK in (b).  }
 \end{figure}

{\it Transition Rate}.--- Now we return to the transient dynamics
and focus on the {\it escape stage} as indicated in Fig.\ 1(c).
This stage is described by a rate process:
\begin{equation}\label{solutions}
\frac{dP_1(t)}{dt}=-\kappa_1P_1+\kappa_2P_2 , \ \
\frac{dP_2(t)}{dt}=-\kappa_2P_2+\kappa_1P_1 ,
\end{equation}
where $\kappa_1$ is the escape rate from SAS to LAS, and
$\kappa_2$ vice versa. 
In what follows, we formulate a way to determine the escape rate,
by contacting~\Eq{solutions} with the numerical simulation. 
First, the solution of~\Eq{solutions} reads
\begin{eqnarray}\label{4}
P_1(t)&=\frac{\kappa_2}{\kappa_1+\kappa_2} +\left[
P_1(0)-\frac{\kappa_2}
{\kappa_1+\kappa_2}\right]\exp^{-(\kappa_1+\kappa_2)t},&
\\\label{5}
P_2(t)&=\frac{\kappa_1}{\kappa_1+\kappa_2}+\left[P_2(0)
-\frac{\kappa_1}
{\kappa_1+\kappa_2}\right]\exp^{-(\kappa_1+\kappa_2)t}.&
\end{eqnarray}
$P_1(0)$ and $P_2(0)$ are certain ``initial" values of the
population probabilities in the escape stage. Rather than $P_1(0)$
and $P_2(0)$, which are not well defined, we use the probabilities
$P_2(t_j)$ at three time points, and simply assume
$t_3-t_2=t_2-t_1=\Delta t$. Then, based on Eqs.\ (\ref{4}) and
(\ref{5}), we obtain
\begin{eqnarray}\label{6}
\kappa_1&=&-\frac{[P_2(t_3)-(K-1)^2P_2(t_1)] \ln(K-1)}{[1-(K-1)^2]
\Delta t} ,
\\\label{7}
\kappa_2&=&-\frac{\ln(K-1)}{\Delta t}-\kappa_1,
\end{eqnarray}
where $K=[P_2(t_3)-P_2(t_1)]/[P_2(t_2)-P_2(t_1)]$.

Below we focus on $\kappa_1$ and formally assume
$\kappa_1=Ce^{-R/\lambda}$. Here $C$ is an irrelevant prefactor,
and the exponential form of $\sim e^{-R/\lambda}$ is associated
with an {\it effective} activation process. In limiting cases,
such as for classical thermal activation, $R$ is the activation
energy and $\lambda$ the temperature; while for quantum tunneling
through a barrier, $R$ is the tunneling action and $\lambda$ the
Plank constant. 
Our present situation is a generalization, i.e.,
quantum-dynamical-tunneling dominated but also thermal-activation
involved. So, we may view $R$ as an effective activation energy
and $\lambda$ an effective Planck constant or temperature.

Physically, we should expect that the transition rate depends on
the driving distance to the critical point $F_c$, i.e.,
$\eta\equiv F^2_c-F^2_0$, since closer to the critical point, more
easily can the transition to the LAS take place. 
Strikingly, Figure 4 displays a perfect scaling behavior for this
dependence. Assuming $R\varpropto\eta^{\alpha}$, our precise
numerical fitting gives $\alpha\simeq 1$. We noticed similar
scaling behavior was found by Dykman~\cite{Dykman011101}, but
where a scaling exponent $\alpha=3/2$ was found instead.

Since our simulation is for a mesoscopic DDO with a bit more than
ten levels involved in the dynamics, we postulate that the scaling
exponent $\alpha=3/2$ is not universal. As in~\cite{Katz07}
and~\cite{low-lying}, our present simulation does not account for
the driving field in the dissipation terms. Although this kind of
treatment is well accepted in vast areas, there is indeed some
counterexamples, for instance, see the most recent
Ref.~\cite{Hon09}. Nevertheless, by transforming the system to a
rotating frame to account for the driving in the dissipation terms
and calculating the fidelity of the SAS, we actually discovered
the same scaling exponent~\cite{Guo}.

We noticed that in Ref.~\cite{Sta06}, scaling behavior of the
transition rate with the driving frequency (but not the driving
strength) was analyzed to give $\alpha\simeq 1.3\sim 1.4$, by a
rough fitting from a few experimental data. 
Meanwhile, in the experiments by Siddiqi {\it et
al.}~\cite{Sid0305}, an effective potential with a barrier height
scaled as $\Delta U^0_{dyn}\propto
\left[1-(F_0/F_c)^2\right]^{3/2}$, was employed to analyze their
measured data by means of the {\it thermal-activation} rate
$\propto \exp\left(-\Delta U^0_{dyn}/k_BT\right) $. 
Based on the same effective potential, a rough WKB analysis should
result in a smaller scaling exponent for {\it quantum} rate
$\propto \exp\left(-\sqrt{\Delta U^0_{dyn}}a/\hbar\right)$, with
$a$ the effective width of the barrier. 
It seems that our above result $\alpha\simeq 1$ is in qualitative
agreement with this analysis. 
Therefore, stronger experimental evidences and more rigorous
theoretical investigations will be helpful to clarify this
interesting issue, particularly with an extension to the
mesoscopic regime as we estimated earlier in this work.

{\it Summary and Discussion}.--- We have investigated the quantum
nonlinear dynamics of the DDO in a mesoscopic regime with a few
more than ten energy levels involved in the dynamics. 
We demonstrated for the first time that the quantum nature
significantly modifies the classical sharp-switching behavior near
the bifurcation point. In terms of Wigner function the state near
the bifurcation point was identified to be a mixture of low- and
high-amplitude states, and the low-amplitude component is nearly a
perfect coherent state. 
More interestingly, near the bifurcation point the transient
dynamics reveals distinct stages of {\it quenching} and {\it
escape}. The latter is well characterized by a rate process, and
the numerically extracted rate displays perfect scaling behavior
with the driving distance to the bifurcation point.

The quantum predictions of this work, in particular the scaling
exponent that differs from the existing result~\cite
{Dykman011101}, raise an open question and deserve further
investigations. 
The statistically mixed nature of the state near the bifurcation
point may also partially explain the discrepancy between the
classical prediction and the measurement data in the JBA
experiment~\cite{Siddiqi027005}. 
Related to this, the mesoscopic DDO may also support quantum weak
measurement in the transient stage, where qubit state can be
updated using a generalized Bayesian rule.

\vspace{1.5cm}
 {\it Acknowledgments.}---
 We acknowledge insightful discussions with Professor Gang Hu.
This work was supported by the National Natural Science Foundation
of China under grants No.\ 10874176, 10875011 and 10575010, the
Major State Basic Research Project under No.\ 2006CB921201, the
973 Program under No.2007CB814805, and the Foundation of Doctoral
Training under No.\ 20060027009.

\end{document}